\documentclass[lettersize,journal]{IEEEtran}
\usepackage{amsmath,amsfonts}
\usepackage{algorithmic}
\usepackage{algorithm}
\usepackage{array}
\usepackage[caption=false,font=normalsize,labelfont=sf,textfont=sf]{subfig}
\usepackage{textcomp}
\usepackage{stfloats}
\usepackage{url}
\usepackage{verbatim}
\usepackage{graphicx}
\usepackage{cite}
\usepackage{breqn}    % 用于自动公式换行
\usepackage{xcolor} 
\usepackage{color}

\usepackage{enumitem}
\usepackage{setspace}

\begin{document}

\title{\huge{Transmit Beamforming Design for ISAC with Stacked Intelligent Metasurfaces}}

\author{ 
{Shunyu~Li,~\IEEEmembership{Student Member,~IEEE}, 
Fan~Zhang,~\IEEEmembership{Student Member,~IEEE},
Tianqi~Mao,~\IEEEmembership{Member,~IEEE}, Rui~Na, Zhaocheng~Wang,~\IEEEmembership{Fellow,~IEEE}, and George K. Karagiannidis,~\IEEEmembership{Fellow,~IEEE}}\vspace{-8mm}

        % <-this % stops a space
\thanks{Copyright (c) 2024 IEEE. Personal use of this material is permitted. However, permission to use this material for any other purposes must be obtained from the IEEE by sending a request to pubs-permissions@ieee.org. }
\thanks{This work was supported in part by National Natural Science Foundation of China under Grant 62401054 and Grant 62088101, in part by the Young Elite Scientists Sponsorship Program by China Association for Science and Technology (CAST) under Grant 2022QNRC001, and in part by National Key R\&D Program of China under Grant 2022YFB3200310. \emph{(Corresponding authors: Tianqi Mao, Rui Na.)}}
\thanks{S. Li, T. Mao and R. Na are with State Key Laboratory of CNS/ATM, Beijing Institute of Technology, Beijing 100081, China. T. Mao is also with MIIT Key Laboratory of Complex-field Intelligent Sensing, Beijing Institute of Technology, Beijing 100081, China. R. Na is also with Yangtze Delta Region Academy of Beijing Institute of Technology (Jiaxing), Jiaxing 314019, China (e-mails: li.shunyu@bit.edu.cn, maotq@bit.edu.cn, narui@bit.edu.cn).}
\thanks{F. Zhang and Z. Wang are with Department of Electronic Engineering, Tsinghua University, Beijing 100084, China (e-mails: zf22@mails.tsinghua.edu.cn, zcwang@tsinghua.edu.cn).}
\thanks{G. K. Karagiannidis is with Department of Electrical and Computer Engineering, Aristotle University of Thessaloniki, Greece (e-mail: geokarag@auth.gr).}
}
\markboth{Journal of \LaTeX\ Class Files,~Vol.~14, No.~8, August~2021}%
{Shell \MakeLowercase{\textit{et al.}}: A Sample Article Using IEEEtran.cls for IEEE Journals}


\maketitle

\begin{abstract}
This paper proposes a transmit beamforming strategy for the integrated sensing and communication (ISAC) systems enabled by the novel stacked intelligent metasurface (SIM) architecture, different from conventional single-layer reconfigurable intelligent surface (RIS) by cascading multiple transmissive metasurface layers, where the base station (BS) simultaneously performs downlink communication and radar target detection via fully passive wave domain beamforming, result in the significant reduction in hardware cost and power consumption.
To ensure superior dual-function performance simultaneously, we design the multi-layer cascading beamformer by maximizing the sum rate of the users while optimally shaping the normalized beam pattern for detection.
A dual-normalized differential gradient descent ($\text{D}^3$) algorithm is further proposed to solve the resulting non-convex multi-objective problem (MOP), where gradient differences and dual normalization are employed to ensure a flexible trade-off between communication and sensing objectives at the gradient level, providing finer control over the optimization process.
Numerical results demonstrate the superiority of the proposed beamforming design in terms of balancing communication and sensing performance.
\end{abstract}

\vspace{-4mm}
\section{Introduction}
Integrated Sensing and Communication (ISAC) is considered one of the most promising technologies for next-generation networks \cite{1,mao,R2_outage_ISAC}.
This approach aims to realize the convergence of communication and sensing systems by sharing hardware platforms, spectrum resources, and even waveforms, which can alleviate spectrum congestion and hardware costs \cite{zhang,R1_fusion_ISAC,R3_secure_ISAC_FD}. Therefore, ISAC techniques can support many emerging applications such as augmented reality, autonomous driving, low-altitude economy, and the Industrial Internet of Everything (IoE), where communication and sensing devices with compact deployment are highly demanded \cite{li2024,5}. 

To further enhance the communication and sensing capabilities, the ISAC framework tends to incorporate the multi-antenna technology for additional spatial degrees of freedom (DoF) achieved by large-scale antenna array \cite{6}.
Despite the fascinating capabilities of digital/hybrid beamforming, classical array-based ISAC systems inevitably suffer from excessive power consumption and hardware cost, resulting from the numerous radio frequency (RF) chains or complex feeding networks with phase shifters and microstrip lines \cite{7,GaoZhen_ISAC_CS}.
To solve this problem, the programmable metasurface, also known as reconfigurable intelligent surface (RIS), can be implemented at the transceiver by replacing the classical phased array antenna, where the RF chains and feeding networks are no longer required \cite{9}. 
%For instance, a 60 GHz programmable dynamic metasurface antenna prototype was introduced in \cite{10}, which directly manipulated the electromagnetic waves by encoding the switch states of the metasurface elements, thus eliminating the need for active phase shifters and amplifiers. 
In \cite{11}, a reconfigurable distributed antenna and reflecting surface aided ISAC system was proposed, combining the distributed gain of the distributed antenna system with the passive beamforming gain of RIS to significantly improve system performance while reducing hardware cost.
Besides, \cite{12} developed a RIS-enabled integrated sensing, communication, and computing system where RF chain-free transmission was realized using RIS as information carriers and modulators. 

The aforementioned literature mainly focused on the single-layer metasurface structure, whose ability to control electromagnetic waves may be limited. 
To this end, the stacked intelligent metasurface (SIM) has recently been proposed as a promising approach which cascades multiple layers of transmissive metasurfaces. 
By performing beamforming directly in the electromagnetic wave domain via passive metasurfaces, SIM significantly reduces hardware complexity and power consumption compared to traditional phased array antennas that rely on numerous active RF components and analog phase shifters \cite{SIMMIMO}.
This SIM technology can further enhance the DoF in manipulating the electromagnetic environment, which exhibits desirable communication performance in terms of multi-user beamforming \cite{13}.
Despite \cite{15} that deployed SIM between the base station (BS) and users/targets for ISAC performance enhancement, research on ISAC system design with the transmit SIM architecture is still in its infancy.

In this context, we propose a transmit beamforming design for ISAC systems equipped with SIM at the BS to perform downlink communication and sensing simultaneously.
Our approach utilizes fully passive wave-domain beamforming, further reducing hardware costs and power consumption by employing equal power allocation across all feeding antennas, in contrast to existing SIM beamforming designs that still rely on individual power control for feeding antennas.
To achieve the desired dual-function performance tradeoff, we optimize the reflection coefficients of SIM elements to form the corresponding sensing beam patterns toward the desired target in the 2D angular domain, while maximizing the total communication data rate.
Furthermore, we propose the Dual-normalized Differential Gradient Descent ($\text{D}^3$) algorithm to tackle this non-convex multi-objective problem (MOP) with coupled variables induced by the cascaded structure of SIM. 
The $\text{D}^3$ algorithm achieves a flexible tradeoff between communication and sensing performance by balancing the gradients of different objective functions. In each iteration, these gradients are first normalized element-wise to the same scale, combined through weighted differences,and then globally normalized within $[-\pi, \pi]$ to comply with the phase shift constraints of the meta-atoms. This approach maintains the physical interpretability of each objective and provides finer control over the optimization process, allowing for flexible steering of the solution toward a desired trade-off.
Simulation results validate the superiority of our proposed transmit beamforming design and elucidate the impact of SIM parameters on dual-function performance, providing useful guidelines for SIM-enabled ISAC implementations.

\begin{figure}[!t]
    \centering
    \includegraphics[width=\linewidth]{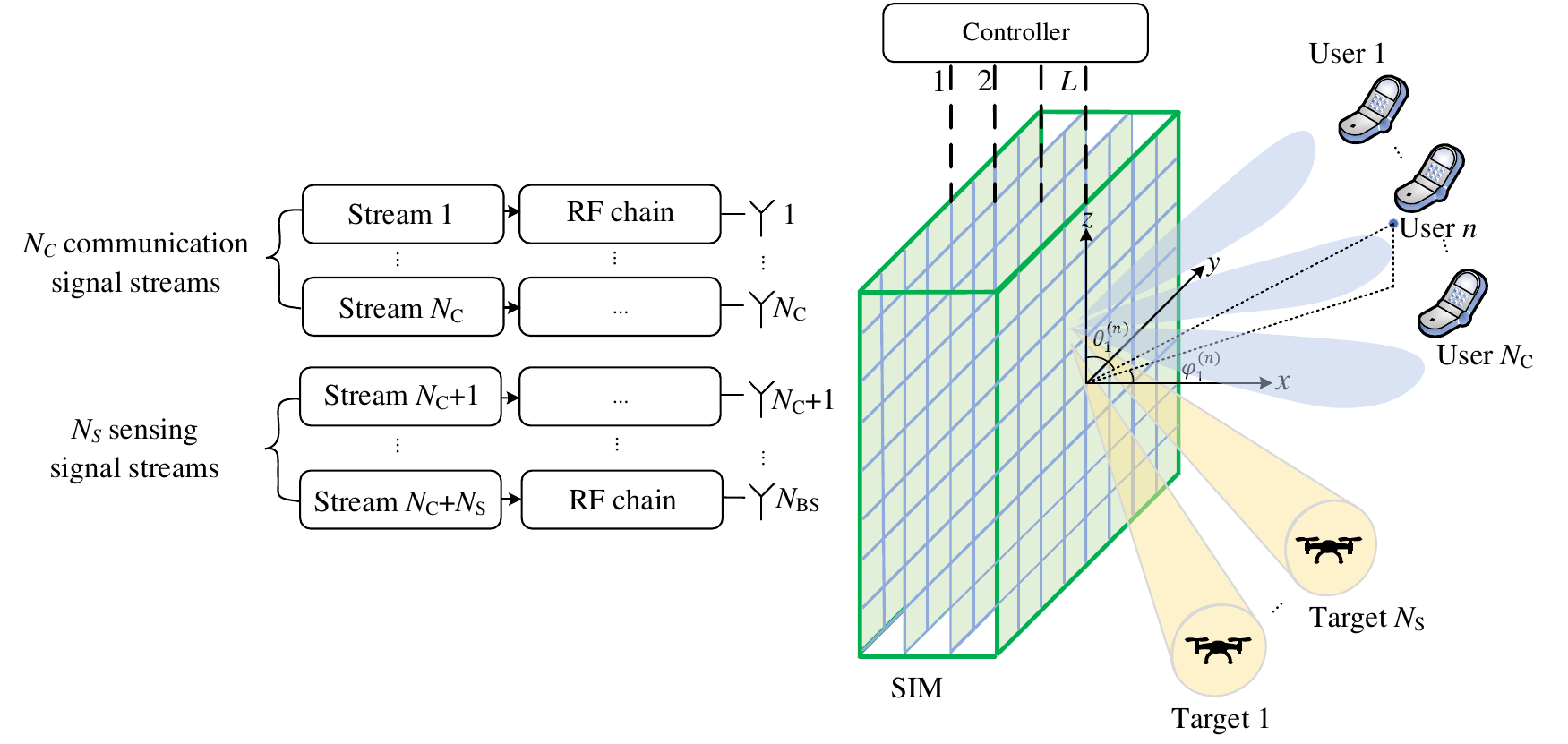}
    \vspace{-3mm}
    \caption{Illustration of the considered SIM-enabled ISAC system. }%Note that \(N_{\mathrm{BS}} \overset{\triangle}{=} N_{\mathrm{C}} + N_{\mathrm{S}}\).
    \label{fig:1}
    \vspace{-5mm}
\end{figure}
\vspace{-3mm}
\section{System Model for SIM-Enabled ISAC}
As shown in Fig.~\ref{fig:1}, we consider a SIM-enabled ISAC system serving \(N_{\mathrm{C}}\) single-antenna users and \(N_{\mathrm{S}}\) sensing targets.
The transmitter (BS) consists of a SIM planar array illuminated by a uniform linear array (ULA) of \(N_{\mathrm{BS}}\) antennas. 
For simplicity and to focus on the SIM-enabled beamforming capabilities, we assume uniform power distribution over the ULA at the BS.
Apart from the $N_{\mathrm{C}}$ parallel communication data streams, we assume $N_{\mathrm{S}}$ independent streams for sensing to extend the DoF for the transmit beamforming design, i.e,  \(N_{\mathrm{BS}} \overset{\triangle}{=} N_{\mathrm{C}} + N_{\mathrm{S}}\), where the number of communication and sensing signal streams is equal to the number of users and targets, respectively \cite{6,7}.

For clarity, the SIM is assumed to have $L$ layers of metasurfaces of size $M = M_{\mathrm{r}} \times M_{\mathrm{c}}$, where $M_{\mathrm{r}}$ and $M_{\mathrm{c}}$ denote the number of meta-atoms in each row and column of each layer, respectively.
The Rayleigh-Sommerfeld diffraction theory provides the mathematical foundation for modeling wave propagation between SIM layers, where each meta-atom acts as a secondary source of spherical waves in accordance with the Huygens-Fresnel principle \cite{SIMMIMO,13,15,16,17}. Based on this theory, the inter-layer channel coefficient \(w_{m, m^{\prime}}^l\) from the \(m^{\prime}\)-th meta-atom on the $(l-1)$-th metasurface layer, i.e., layer $(l-1)$, to the $m$-th meta-atom on layer $l$, is expressed by
\begin{equation}
\label{eq:1}
w_{m, m^{\prime}}^l=\frac{A_{\text{t}} \cos \chi_{m, m^{\prime}}^l}{r_{m, m^{\prime}}^l}\left(\frac{1}{2 \pi r_{m, m^{\prime}}^l}-j \frac{1}{\lambda}\right) e^{j 2 \pi \frac{r_{m, m^{\prime}}^l}{\lambda}},
\end{equation}
where \(\lambda\) is the wavelength, $A_{\text{t}}$ represents the size of each meta-atom, \(r_{m, m^{\prime}}^l\) denotes the transmission distance, while \(\chi_{m, m^{\prime}}^l\) specifies the angle between the propagation direction and the normal direction of the layer $(l-1)$.
By applying \eqref{eq:1} to each meta-atom, we obtain the inter-layer diffraction matrix, denoted as \(\mathbf{W}^l \in \mathbb{C}^{M \times M}\) for \(l =2,3,\cdots,L\).
Besides, we have \(\mathbf{W}^1 \in \mathbb{C}^{M \times N_{\mathrm{BS}}}\) for \(l=1\), which represents the diffraction matrix from the feeding antenna array to the layer $1$ of the SIM.
Additionally, the diagonal phase shift matrix \(\mathbf{\Phi}^l\) of the layer $l$ of the SIM can be formulated as
\begin{equation}\
\label{eq:2}
\mathbf{\Phi}^l=\operatorname{diag}\left(e^{j \theta_1^l}, e^{j \theta_2^l}, \cdots,e^{j \theta_m^l},\cdots, e^{j \theta_M^l}\right),
\end{equation}
where \(e^{j \theta_m^l}\) denotes the phase shift applied by the \(m\)-th meta-atom on layer $l$ with $\theta_m^l \in[0,2 \pi)$, for \(l =1,2,\cdots,L\) and \(m =1,2,\cdots,M\).
Afterwards, the SIM-enabled beamforming matrix \(\mathbf{F}_{\mathrm{SIM}}\in \mathbb{C}^{M \times N_{\mathrm{BS}}}\) can be expressed as
\begin{equation}
\label{eq:3}
\mathbf{F}_{\mathrm{SIM}}=\mathbf{\Phi}^L \mathbf{W}^L \mathbf{\Phi}^{L-1} \mathbf{W}^{L-1}\cdots\mathbf{\Phi}^1 \mathbf{W}^1 .
\end{equation}

By adopting the Saleh-Valenzuela channel model for MIMO systems \cite{7}, the channel between the SIM and the \(n\)-th communication user, i.e., user $n$, for \(n=1,2,\ldots,N_{\mathrm{C}}\) can be characterized as 
\begin{equation}
\label{eq:4}
\mathbf{h}_n=\sqrt{\frac{M}{Q_n}} \sum_{q=1}^{Q_n} g_q^{(n)} \boldsymbol{\alpha}^{H}\left( \theta_q^{(n)},\varphi_q^{(n)}\right), 
\end{equation}
where \( Q_n \) is the number of resolvable channel paths, while \( g_q^{(n)} \) represents the channel gain for \(q=1,2,\cdots,Q_n\). 
Specifically, for the line-of-sight (LoS) path, the channel gain is distributed as $g_{q=1}^{(n)} \sim \mathcal{CN}(0, \zeta)$, while for non-line-of-sight (NLoS) paths, the channel gain is distributed as $g_{q>1}^{(n)} \sim \mathcal{CN}(0, 0.01\zeta)$ \cite{7,13}.
Here, $\zeta$ denotes the distance-dependent path loss modeled as $\zeta = C_0 d_n^{-\alpha}$, where $C_0$ is the free space path loss, $\alpha$ is the path loss exponent and $d_n$ is the distance from the SIM to the user $n$.
Moreover, \(\theta_q^{(n)}\) and \(\varphi_q^{(n)}\) represent the elevation and azimuth angles of departure (AoD) of the \(q\)-th channel path.
%relative to those of the LoS path component 
\( \boldsymbol{\alpha}\left( \theta, \varphi \right) \in \mathbb{C}^{M \times 1} \) denotes the channel steering vector, which is a function of elevation and azimuth angles expressed by
\begin{dmath}
\label{eq:5}
\boldsymbol{\alpha}(\theta, \varphi)=\left[1, \cdots, e^{-j 2 \pi\left(M_{\mathrm{r}}-1\right) \sin (\theta) \cos (\varphi) \frac{d_y}{\lambda}}\right] \otimes\left[1, \cdots, e^{-j 2 \pi\left(M_{\mathrm{c}}-1\right) \sin (\theta) \sin (\varphi) \frac{d_z}{\lambda}}\right],
\end{dmath}
where \(\otimes\) stands for the Kronecker product, and \(d_y\) and \(d_z\) represent the horizontal and perpendicular spacings between adjacent meta-atoms, respectively \cite{18}.

The transmit signal at the BS before passing through the SIM is denoted as $\boldsymbol{x} \in \mathbb{C}^{N_{\mathrm{BS}} \times1}$, where $\mathrm{E}\{\boldsymbol{x}\}=\mathbf{0}$ and $\mathrm{E}\left\{\boldsymbol{x} \boldsymbol{x}^{\mathrm{H}}\right\}=\boldsymbol{I}_{N_{\mathrm{BS}}}$. The first $N_{\mathrm{C}}$ elements of $\boldsymbol{x}$, i.e., $x_1, x_2, \cdots, x_{N_{\mathrm{C}}}$, represent the information symbols for the $N_{\mathrm{C}}$ communication users, while the remaining $N_{\mathrm{S}}$ elements, i.e., $x_{N_{\mathrm{C}}+1}, \cdots, x_{N_{\mathrm{BS}}}$, correspond to the sensing waveforms for the $N_{\mathrm{S}}$ sensing targets.
Hence, the received signals by different users, denoted as \(\boldsymbol{y} \in \mathbb{C}^{N_{\mathrm{C}} \times 1}\), can be expressed as 
\begin{equation}
\label{eq:6}
\boldsymbol{y}=\mathbf{H} \mathbf{F}_{\mathrm{SIM}} \boldsymbol{x}+\boldsymbol{n},
\end{equation}
where \(\mathbf{H} \overset{\triangle}{=} \left[\mathbf{h}_1, \mathbf{h}_2, \cdots, \mathbf{h}_{N_{\mathrm{C}}}\right]^T \in \mathbb{C}^{N_{\mathrm{C}} \times M}\) denotes the total channel matrix, and \(\boldsymbol{n} \in \mathbb{C}^{N_{\mathrm{C}} \times 1}\) is the additive white Gaussian noise (AWGN) vector with \(\boldsymbol{n} \sim \mathcal{C} \mathcal{N}\left(\mathbf{0}, \sigma^2 \mathbf{I}_{N_{\mathrm{C}}}\right)\). Here, \(\sigma^2\) is the noise power at the receivers, and \(\mathbf{I}_{N_{\mathrm{C}}}\) is the \(N_{\mathrm{C}}\)-by-\(N_{\mathrm{C}}\) identity matrix.
For user $n$, both the signals from the other users and sensing targets are regarded as interference. Therefore, the signal-to-interference-plus-noise ratio (SINR) of user $n$ can be expressed as
\begin{equation}
\label{eq:7}
\gamma_n=\frac{\left|\left[\mathbf{H F}_{\mathrm{SIM}}\right]_{n, n}\right|^2}{\sum_{i=1, i \neq n}^{N_{\mathrm{BS}}}\left|\left[\mathbf{H F}_{\mathrm{SIM}}\right]_{n, i}\right|^2+\sigma^2},
\end{equation}
where \(\left[\mathbf{H F}_{\mathrm{SIM}}\right]_{n, i}\) denotes the element in the \(n\)-th row and \(i\)-th column of the \(\mathbf{H F}_{\mathrm{SIM}}\). 
Then, the sum rate\footnote{The sum rate here strictly refers to the achievable spectral efficiency measured in bits per second per Hz (bits/s/Hz).} of the \(N_{\mathrm{C}}\) users, according to the Shannon-Hartly theorem, 
can be written as
\begin{equation}
\label{eq:8}
R_{\text{sum}} = \sum_{n=1}^{N_{\mathrm{C}}} \log_2 \left(1 + \gamma_n\right).
\end{equation}

In terms of sensing, communication signals are regarded as supplementary to enhance the sensing signals. Let \(\{\psi_1, \psi_2, \cdots, \psi_{N_{\mathrm{D}}}\}\) and \(\{\phi_1, \phi_2, \cdots, \phi_{N_{\mathrm{D}}}\}\) denote the sampling points evenly distributed in the elevation and azimuth angle domains, respectively. Then the beam pattern gain \(\mathbf{P}_{\mathrm{S}}\in \mathbb{R}^{N_{\mathrm{D}} \times N_{\mathrm{D}}}\) of the SIM in the direction  \(\{\psi_j, \phi_k\}\) for \(j, k = 1,2, \cdots, N_{\mathrm{D}}\) can be expressed as
\begin{equation}
\label{eq:9}
\left[\mathbf{P}_{\mathrm{S}}\right]_{j, k}=\boldsymbol{\alpha}^H\left(\psi_j, \phi_k\right) \mathbf{F}_{\mathrm{SIM}} \mathbf{F}_{\mathrm{SIM}}^H \boldsymbol{\alpha}\left(\psi_j, \phi_k\right),
\end{equation}
where \(\boldsymbol{\alpha}\left(\psi_j, \phi_k\right)\) is the channel steering vector obtained by \eqref{eq:5}. Then the normalized beam pattern  $\overline{\mathbf{P}}_{\mathrm{S}}$ is calculated as
\begin{equation}
\label{eq:10}
\overline{\mathbf{P}}_{\mathrm{S}} = {\mathbf{P}_{\mathrm{S}}}\Big/{\|\mathbf{P}_{\mathrm{S}}\|_1}.
\end{equation}
Given the desired beam pattern \(\mathbf{P}_{\mathrm{D}} \in \mathbb{R}^{N_{\mathrm{D}} \times N_{\mathrm{D}}}\), we define the mean square error (MSE) between \(\mathbf{P}_{\mathrm{D}}\) and \(\overline{\mathbf{P}}_{\mathrm{S}}\) as the beam-matching error \cite{6,7}, calculated as
\begin{equation}
\label{eq:11}
J_{\text{MSE}} = \|\overline{\mathbf{P}}_{\mathrm{S}} - \mathbf{P}_{\mathrm{D}}\|_2^2.
\end{equation}
Here, \(\|\cdot\|_1\) and \(\|\cdot\|_2\) denote the \(\ell_1\) and \(\ell_2\) norms, respectively.
\vspace{-3mm}
\section{SIM-Based Transmit Beamforming Design}
\subsection{Problem Fomulation}
To facilitate downlink communication and sensing with desirable performance trade-off, an MOP is formulated to maximize the user sum rate and concurrently minimize the beam-matching error under uniform transmit power allocation. These can be realized by configuring the phase shifts imposed by the SIM meta-atoms.
By defining \(\boldsymbol{\vartheta} \overset{\triangle}{=} \{\boldsymbol{\theta}^1, \boldsymbol{\theta}^2, \cdots, \boldsymbol{\theta}^L\}\) with \(\boldsymbol{\theta}^l \overset{\triangle}{=} \left[\theta_1^l, \theta_2^l, \cdots, \theta_M^l\right]^T\), the MOP is shown as
\begin{subequations}
\label{eq:12}
\begin{alignat}{2}
&\text{P1}: 
 &&\max_{\boldsymbol{\vartheta}} \quad  R_{\text{sum}} \label{eq:12a} \\
& \quad\quad\:&&\min_{\boldsymbol{\vartheta}} \quad  J_{\text{MSE}} \label{eq:12b} \\
& \text{s.t.}  && \begin{aligned}[t]
 & \theta_m^l \in [0, 2\pi), 
 & \forall l = 1,2,\cdots,L,\forall m = 1,2,\cdots,M.
\end{aligned} \label{eq:12c}
\end{alignat}
\end{subequations}

$(\text{P1})$ is inherently non-convex due to the form of the objective function in \eqref{eq:12a} and \eqref{eq:12b}. Furthermore, the strong coupling among the multiple phase shift matrices across the $L$ layers of the SIM makes the problem even more intractable. 
\subsection{$\text{D}^3$ Algorithm}
%The proposed $\text{D}^3$ algorithm is an adaptation of the traditional gradient descent method used to optimize the phase shift matrix of SIM \cite{13,14}. 
To address the aforementioned problem, we propose an efficient Dual-Normalized Differential Gradient Descent ($\text{D}^3$) algorithm to achieve a quasi-optimal solution. More specifically, our approach addresses a MOP that includes both communication and sensing tasks. It includes gradient differences and additional normalization steps that adjust the gradient components to similar scales across different tasks. This dual-normalization ensures that the optimization process achieves a balanced trade-off between communication and sensing tasks, and avoids over-promoting any single objective.

First, the phase shifts $\theta_m^l \in [0, 2\pi)$ for $l = 1, 2, \cdots, L$ and $m = 1, 2, \cdots, M$ are initialized as, e.g., a uniform random distribution. 
Next, the partial derivatives of $J_{\text{MSE}}$ in \eqref{eq:12b} with respect to $\theta_m^l$ are derived as 
\begin{equation}
\label{eq:13}
\frac{\partial J_{\text{MSE}}}{\partial \theta_m^l} = \frac{1}{{N_{\mathrm{D}}}^2} \sum_{j=1}^{N_{\mathrm{D}}} \sum_{k=1}^{N_{\mathrm{D}}} 2\left([\overline{\mathbf{P}}_{\mathrm{S}}]_{j, k}-\left[\mathbf{P}_{\mathrm{D}}\right]_{j, k}\right) \cdot [\overline{\mathbf{E}}]_{j, k},
\end{equation}
where $\overline{\mathbf{E}}$ represents the partial derivatives of $\overline{\mathbf{P}}_{\mathrm{S}}$ in terms of phase shifts $\theta_m^l$, expressed as
\begin{equation}
\label{eq:14}
[\overline{\mathbf{E}}]_{j, k} = \frac{\left\|\mathbf{P}_{\mathrm{S}}\right\|_1 [\mathbf{E}]_{j, k} - \left[\mathbf{P}_{\mathrm{S}}\right]_{j, k} \|\mathbf{E}\|_1}{\left\|\mathbf{P}_{\mathrm{S}}\right\|_1^2}.
\end{equation}
Furthermore, $\mathbf{E}$ denotes the partial derivatives of $\mathbf{P}_{\mathrm{S}}$ with respect to phase shifts $\theta_m^l$, which can be calculated as
\begin{equation}
\label{eq:15}
[\mathbf{E}]_{j, k} = 2\operatorname{Im}\{ e^{j \theta_m^l}  \boldsymbol{\alpha}\left(\psi_j, \phi_k\right)^H \mathbf{V}_{:, m}^l 
\mathbf{U}_{m,:}^l \mathbf{F}_{\text{SIM}}^H \boldsymbol{\alpha}\left(\psi_j, \phi_k\right) \}.
\end{equation}
$\operatorname{Im}\left\{ \cdot \right\}$ denotes the imaginary part, and $\mathbf{V}_{:, m}^l$ and $\mathbf{U}_{m,:}^l$ denote the $m$-th column of $\mathbf{V}^l$ and $m$-th row of $\mathbf{U}^l$, defined by
\begin{align}
\label{eq:16}
\mathbf{U}^l &\overset{\triangle}{=} \begin{cases}
\mathbf{W}^l \boldsymbol{\Phi}^{l-1} \mathbf{W}^{l-1} \cdots \boldsymbol{\Phi}^2 \mathbf{W}^2 \boldsymbol{\Phi}^1 \mathbf{W}^1, &\:\:\:\:\: \text { if } l \neq 1, \\ 
\mathbf{W}^1, &\:\:\:\:\: \text { if } l=1,
\end{cases}\\
\label{eq:17}
\mathbf{V}^l &\overset{\triangle}{=} \begin{cases}
\boldsymbol{\Phi}^L \mathbf{W}^L \boldsymbol{\Phi}^{L-1} \mathbf{W}^{L-1} \cdots \boldsymbol{\Phi}^{l+1} \mathbf{W}^{l+1}, & \text { if } l \neq L, \\ 
\mathbf{I}_M, & \text { if } l=L.
\end{cases}
\end{align}

Besides, the partial derivatives of $R_{\text{sum}}$ in \eqref{eq:12a} with respect to $\theta_m^l$ are derived as 
\begin{equation}
\label{eq:18}
\frac{\partial R_{\text{sum}}}{\partial \theta_m^l} = 2 \log_2 e \sum_{p=1}^{N_{\mathrm{C}}} \delta_p\left(\eta_{p, p}-\gamma_p \sum_{q=1, q \neq p}^{N_{\mathrm{BS}}} \eta_{p, q}\right),
\end{equation}
where $\delta_p$ and $\eta_{p, q}$ in \eqref{eq:18} are given by \cite{13}
\begin{align}
\label{eq:19}
\delta_p &= \frac{1}{\sum_{q=1}^{N_{\mathrm{C}}}\left|\left[\mathbf{HF}_{\mathrm{SIM}}\right]_{p, q}\right|^2 + \sigma^2},\\
\label{eq:20}
\eta_{p, q} &= \operatorname{Im}\left\{ \left[\mathbf{H V}^l\right]_{p, m} \left[\mathbf{U}^l\right]_{m, q} \left[\mathbf{H F}_{\text{SIM}}\right]_{p, q}^* e^{j \theta_m^l} \right\}.
\end{align}

%\begin{algorithm}[!t]
%\caption{Proposed $\text{D}^3$ Algorithm to Solve $(P 2)$}
\label{alg1}

However, the partial derivatives obtained from \eqref{eq:13} and \eqref{eq:18} exhibit different levels of values due to their distinct physical interpretations. 
To ensure that both communication and sensing gradients are fairly considered in the optimization process, we first normalize both gradients in an element-wise manner, expressed as
\begin{align}
\tilde{g}_{\text{sens}} &= \frac{\frac{\partial J_{\text{MSE}}}{\partial \theta_m^l}}{\sqrt{\left(\frac{\partial J_{\text{MSE}}}{\partial \theta_m^l}\right)^2+\epsilon}}, \label{eq:21} \\ 
\tilde{g}_{\text{com}} &= \frac{\frac{\partial R_{\text{sum}}}{\partial \theta_m^l}}{\sqrt{\left(\frac{\partial R_{\text{sum}}}{\partial \theta_m^l}\right)^2+\epsilon}}, \label{eq:22}
\end{align}
where $\epsilon$ is the smoothing term as a small constant for numerical stability.
Here, $\tilde{g}_{\text{sens}}$ and $\tilde{g}_{\text{com}}$ denote the first normalized gradients of sensing and communication objectives, respectively, whose magnitudes are scaled to the same range of $(-1, 1)$.
The differential gradient \(\mathbf{G} \in \mathbb{C}^{M \times L}\) is then formed by a weighted difference of $\tilde{g}_{\text{sens}}$ and $\tilde{g}_{\text{com}}$, expressed as
\begin{equation}
\label{eq:23}
[\mathbf{G}]_{m, l}=w_1 \tilde{g}_{\text {sens}}-w_2 \tilde{g}_{\text {com}},
\end{equation}
where $w_1$ and $w_2$ are weighting coefficients for the sensing and communication objectives within the interval of $[0, 1]$, respectively.
In every iteration, the relative impacts of each objective on $[\mathbf{G}]_{m, l}$ are proportional to the adjustable weights via \eqref{eq:23}, thus ensuring a flexible trade-off with desirable dual-functional performances for various application requirements.

Additionally, in order to mitigate gradient explosion and vanishing issues during optimization \cite{16}, global normalization is applied to the differential gradient. Thus, the dual-normalized differential gradient \(\overline{\mathbf{G}}\) can be expressed as
\begin{equation}
\label{eq:24}
[\overline{\mathbf{G}}]_{m, l} = \frac{\pi}{\max (\mathbf{G})} \cdot [\mathbf{G}]_{m, l},  
\end{equation}
where $m = 1, 2, \cdots, M$, $ \ l = 1, 2, \cdots, L$, and \(\max (\cdot)\) denotes the operation of extracting the largest element. 

Ultimately, the phase shifts \(\theta_m^l \in \boldsymbol{\vartheta}\) across each meta-atom in SIM can be updated via \(\overline{\mathbf{G}}\) at each iteration, expressed as
\begin{equation}
\label{eq:25}
\theta_m^l \leftarrow \theta_m^l-\mu \cdot[\overline{\mathbf{G}}]_{m, l},
\end{equation}
where $\mu$ is the step size. Specifically, we adopt an exponentially decreasing learning rate schedule as the iteration proceeds, which is updated by %\cite{16}
\begin{equation}
\label{eq:26}
\mu \leftarrow \mu \beta,
\end{equation}
where \(\beta\) is a hyperparameter determining the decay rate, satisfying \(0 < \beta < 1\).

By iteratively applying \eqref{eq:13}-\eqref{eq:26}, the objective function \eqref{eq:12a} and \eqref{eq:12b} will reach convergence when its decrease is smaller than a preset threshold or reaching the maximum iteration number.
%For clarity, the detailed procedures of the proposed $\text{D}^3$ algorithm are summarized in Algorithm \ref{alg1}.

{\subsection{The Algorithm Complexity Analysis}}
Next, we evaluate the computational complexity of the proposed algorithm. 
Specifically, the complexity of $\text{D}^3$ algorithm is $\mathcal{O}(I_{\text{D}^3}(N_{\mathrm{D}}^2 L M^2+L^2M^3))$, where $I_{\text{D}^3}$ denotes the number of iterations. 
Also $\mathcal{O}\left(L M^2 N_{\mathrm{BS}}\right)$,
$\mathcal{O}\left(L^2M^3\right)$,
$\mathcal{O}\left(N_{\mathrm{D}}^2 L M^2\right)$,
$\mathcal{O}\left(M L\right)$,
$\mathcal{O}\left(1\right)$,
denote computational complexity of the
calculation for beamforming matrix \eqref{eq:3},
auxiliary matrices \eqref{eq:17},
gradient of sensing objctive \eqref{eq:13}-\eqref{eq:15}, 
gradient of communication objctive \eqref{eq:18}-\eqref{eq:20}, 
dual-normalized gradient difference \eqref{eq:21}-\eqref{eq:25}, 
and step size update \eqref{eq:26}, respectively.

\section{Numerical Results}
This section validates the effectiveness of the transmit beamforming design with the proposed $\text{D}^3$ algorithm. The simulation parameters are summarized in Table~\ref{tab:simulation_parameters}.
The SIM is oriented parallel to the $y$-$z$ plane and is centered along the $x$-axis, with the middle of the outermost layer situated at the origin of the Cartesian coordinate system while the innermost layer is found at the plane $x=-0.05 \, \text{m}$. 
For simplicity, we consider a square metasurface structure with $M_{\mathrm{r}}$ = $M_{\mathrm{c}}$. Furthermore, we assume half-wavelength spacing between adjacent feeding antennas at the BS and between adjacent meta-atoms at each metasurface layer. The communication users are assumed to be distributed on the plane $x = 10 \,\text{m}$.
% For the $n$-th user, the azimuth angle $\varphi_1^{(n)}$ is defined as the angle between the projection of the LOS path onto the $x$-$y$ plane and the positive $x$-axis, while the elevation angle $\theta_1^{(n)}$ is defined as the angle between the LOS path and the positive $z$-axis. Consequently, the distance between the $n$-th user and the SIM is determined as $d_n = \frac{10}{\sin(\theta_1^{(n)}) \cos(\varphi_1^{(n)})}$, for $n = 1, \ldots, N_{\mathrm{C}}$. 
For the NLoS paths of each user, a random angular spread of $10^\circ$ around the LoS AoDs is introduced, where $\theta_{2,3}^{(n)} \sim {U(\theta_{1}^{(n)}-5^\circ,\theta_{1}^{(n)}+5^\circ)}$ and $\varphi_{2,3}^{(n)} \sim {U(\varphi_{1}^{(n)}-5^\circ,\varphi_{1}^{(n)}+5^\circ)}$.
The desired beam pattern for sensing is assumed as
\begin{equation}
\label{eq:27}
\left[\mathbf{P}_{\mathrm{D}}\right]_{j,k} =
\begin{cases}
1 & \text{if } (j,k) \in \{(9, 27), (27, 9)\}, \\
0 & \text{otherwise},
\end{cases}
\end{equation}
where \(j, k = 1,2, \cdots, N_{\mathrm{D}}\) correspond to specific elevation and azimuth angles. Specifically, indices 9 and 27 correspond to the angle ranges \([-45^\circ, -40^\circ]\) and \([45^\circ, 50^\circ]\), respectively.

\begin{table}[t]
\centering
\caption{Simulation Parameters}
\begin{tabular}{|c|c|}
\hline
\textbf{Parameter} & \textbf{Value} \\
\hline
Carrier frequency $f_c$ (GHz) & $28$ \\
\hline
% Wavelength $\lambda$ (mm) & $10.7$ \\
% \hline
Thickness of the SIM (m) & $0.05$ \\
\hline
Inter-layer spacing (m) & $0.05/L$ \\
\hline
Meta-atom spacing $d_y, d_z$ & $\lambda/2$ \\
\hline
Meta-atom area $A_{\text{t}}$ & $\lambda^2/4$ \\
\hline
Total transmission power (dBm) & $20$ \\
\hline
BS antenna gain (dBi) & $5$ \\
\hline
User antenna gain (dBi)& $0$ \\
\hline
Path loss constant $C_0$ (dB) & $-32$ \\
\hline
Path loss exponent $\alpha$ & $3.5$ \\
\hline
AWGN noise power $\sigma^2$ (dBm) & $-104$ \\
\hline
Number of users $N_{\mathrm{C}}$ & $4$ \\
\hline
Number of targets $N_{\mathrm{S}}$ & $2$ \\
\hline
Number of resolvable channel paths $Q_n$ & $3$ \\
\hline
Number of sample points $N_{\mathrm{D}}$ & $36$ \\
\hline
User coordinates (m) & $x = 10$ \\
\hline
User 1 AoD $(\theta_1^{(1)}, \varphi_1^{(1)})$  & $(60^{\circ}, 45^{\circ})$ \\
\hline
User 2 AoD $(\theta_1^{(2)}, \varphi_1^{(2)})$  & $(60^{\circ}, 35^{\circ})$ \\
\hline
User 3 AoD $(\theta_1^{(3)}, \varphi_1^{(3)})$  & $(-60^{\circ}, -45^{\circ})$ \\
\hline
User 4 AoD $(\theta_1^{(4)}, \varphi_1^{(4)})$  & $(-60^{\circ}, -30^{\circ})$ \\
\hline
Smoothing term $\epsilon$  & $ 10^{-8}$ \\
\hline
\end{tabular}
\label{tab:simulation_parameters}
%\vspace{-5mm}
\end{table}
\begin{figure}[t]
    \centering
    \includegraphics[width=0.89\columnwidth]{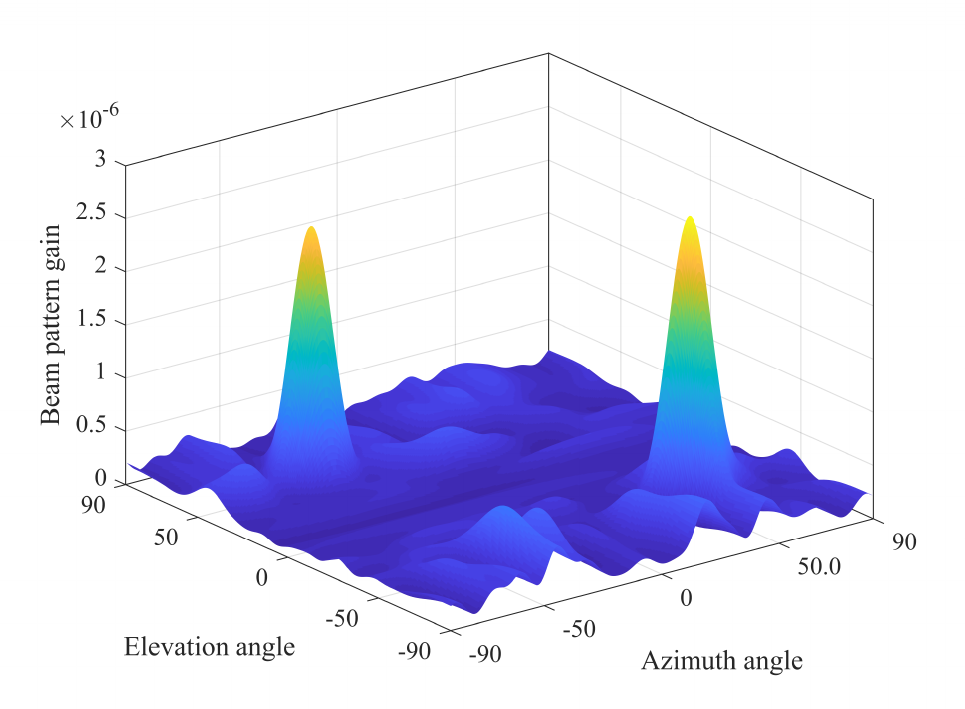}
    \vspace{-3mm}
    \caption{Transmit beam patterns in the angle space with the proposed beamforming design.}
    \label{fig:2}
    \vspace{-5mm}
\end{figure} 

To prevent premature convergence to an undesirable local optimum, five distinct sets of initial phase shifts are generated. The $\text{D}^3$ algorithm is then executed in parallel for each of these sets \cite{13}, \cite{16}. 
Following the parallel execution, the quasi-optimal solution for $(\text{P1})$ is selected based on the values of $w_1$ and $w_2$.
Specifically, we select the solution with the lowest $J_{\text{MSE}}$ when $w_1 > w_2$, prioritizing sensing performance, or the highest $R_{\text{sum}}$ when $w_1 \leq w_2$, favoring communication performance.
The algorithm iterates until either the relative change in the objective function falls below a threshold of $10^{-6}$ or the maximum number of 60 iterations is reached. Unless otherwise specified, we set the initial learning rate to $\eta = 1$ and the decay parameter to $\beta = 0.5$.
All simulation results
%,with the exception of those presented in Figure~\ref{fig:2}, 
are obtained by averaging $100$ independent experiments.

Figure~\ref{fig:2} shows the beam pattern obtained by the proposed beamforming design using the $\text{D}^3$ algorithm, where \( M = 100 \), \( L = 7 \), and \( w_1 = w_2 = 1 \).
It can be observed that the beam pattern gain reaches its maximum (about $2.5 \times 10^{-6}$) in the directions towards the sensing targets within the regions of $[-45^\circ, -40^\circ]$ in elevation angle space and $[45^\circ, 50^\circ]$ in azimuth angle space, as well as $[45^\circ, 50^\circ]$ in elevation angle space and $[-45^\circ, -40^\circ]$ in azimuth angle space, where the beam-matching error is calculated as $J_{\text{MSE}} = -12.79 \,$dB. Meanwhile, the beam peaks for the four communication users are located at  $(-60^\circ, -45^\circ)$, and $(-60^\circ, -35^\circ)$. Although the average strength of the communication beams is not comparable to the sensing beams, i.e., below the gain of $0.75 \times 10^{-6}$, superior data throughput with $R_{\text{sum}} = 15$ bit/s/Hz can be achieved with the proposed $\text{D}^3$ algorithm. Therefore, the proposed beamforming design successfully realizes a desirable balance between sensing and communication functions.

%When $w_1=1$ and $w_2=0$, the beamforming design functions as a sensing-only scheme, while $w_1=0$ and $w_2=1$ conversely represents a communication-only scheme. For cases where both $w_1$ and $w_2$ are non-zero, the system functions as an ISAC scheme, with their ratio determining the priority of sensing or communication in the system.

%Furthermore, Fig.~\ref{fig:3} shows that the sum rate of users increases with the number of meta-atoms $M$ and the number of layers $L$. This improvement is due to the additional flexibility in shaping the electromagnetic wavefront provided by the multi-layer massive metasurface array, which can enable more sophisticated beamforming patterns to better accommodate the dual-functional requirements.

\begin{figure}[t]
    \centering
    \includegraphics[width=0.85\columnwidth]{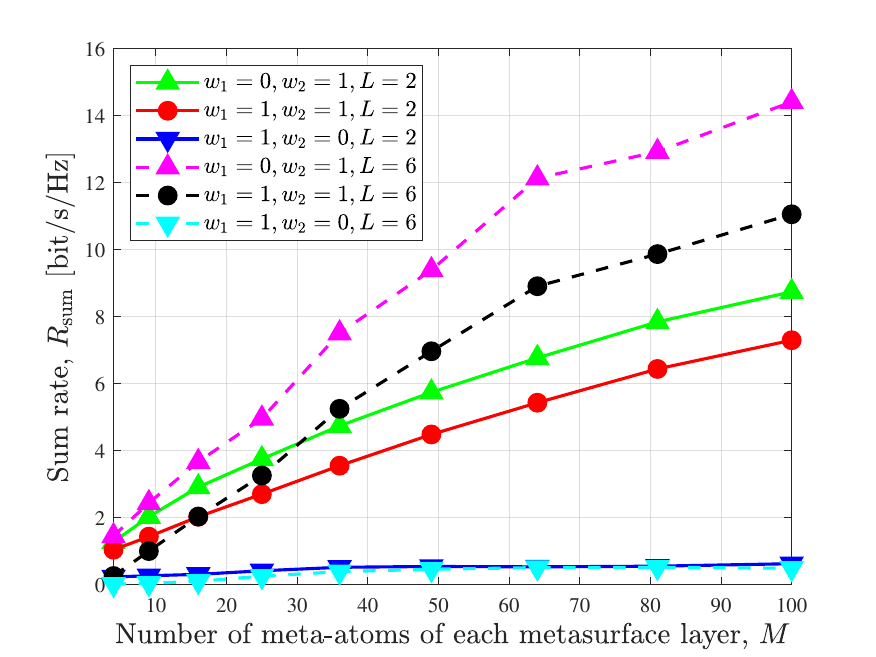}
    \vspace{-3mm}
    \caption{Sum rate of communication users versus number of meta-atoms.}
    \label{fig:3}
    \vspace{-5mm}
\end{figure}

Figure~\ref{fig:3} illustrates the communication performance of the proposed beamforming design with respect to the number of meta-atoms $M$ under different system parameters. It can be observed that the sensing-only scheme ($w_1=1$, $w_2=0$) exhibits poor communication performance, with sum rates that are almost nil. In contrast, the ISAC scheme with $w_1=w_2=1$ achieves a sum rate approaching that of the communication-only scheme ($w_1=0$, $w_2=1$), while simultaneously achieving a beam-matching error below $J_{\text{MSE}} = -12.7 \,$dB for $L=2$, $M=100$ and $-12.8 \,$dB for $L=6$, $M=100$. This validates that the $\text{D}^3$ algorithm can achieve both desirable communication and sensing capabilities concurrently.

\begin{figure*}[!th]
\vspace*{-2mm}

\begin{minipage}[t]{0.48\linewidth}
    \centering
    \includegraphics[width=\linewidth, keepaspectratio]{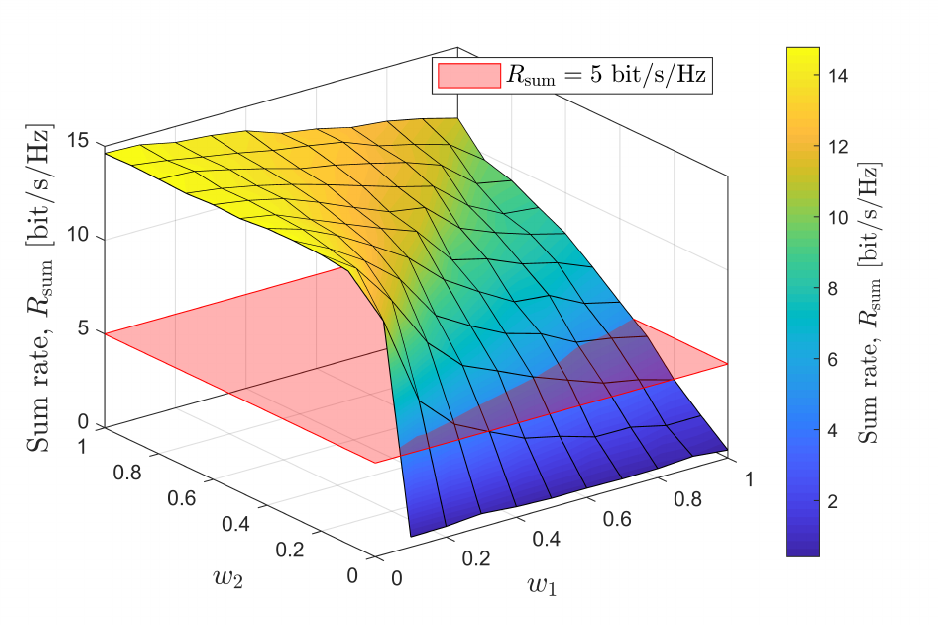}
    \vspace*{-4mm}
    \caption{Sum rate $R_{\text{sum}}$ versus $w_1$ and $w_2$ ($M = 100$, $L = 6$).}
    \label{fig:4}
\end{minipage}
\hfill
\begin{minipage}[t]{0.48\linewidth}
    \centering
    \includegraphics[width=\linewidth, keepaspectratio]{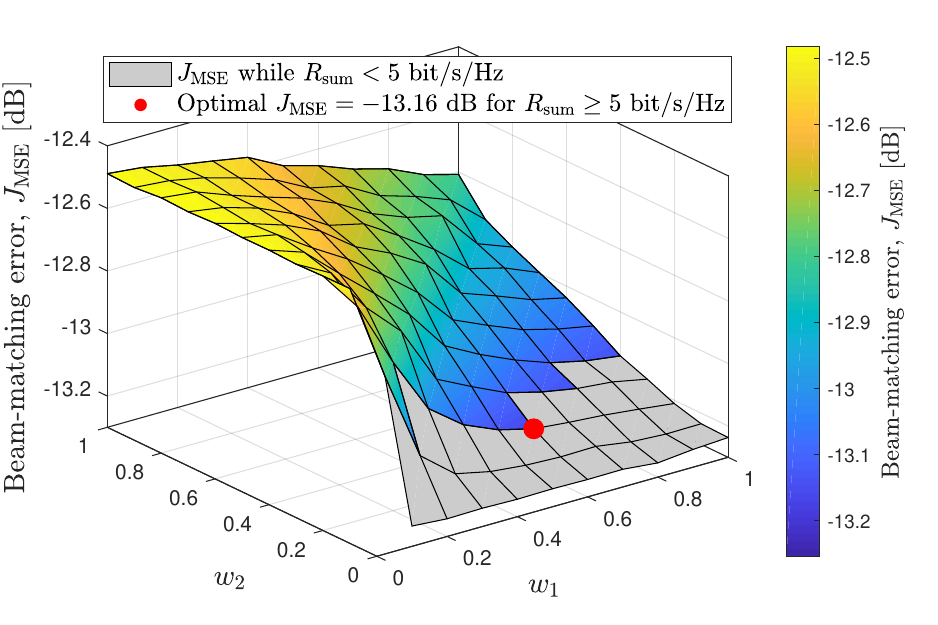}
    \vspace*{-4mm}
    \caption{Beam-matching error $J_{\text{MSE}}$ versus $w_1$ and $w_2$ ($M = 100$, $L = 6$).}
    \label{fig:5}
\end{minipage}

\vspace*{-7mm}
\end{figure*}

\begin{figure}[t]
    \centering
    \includegraphics[width=0.8\columnwidth]{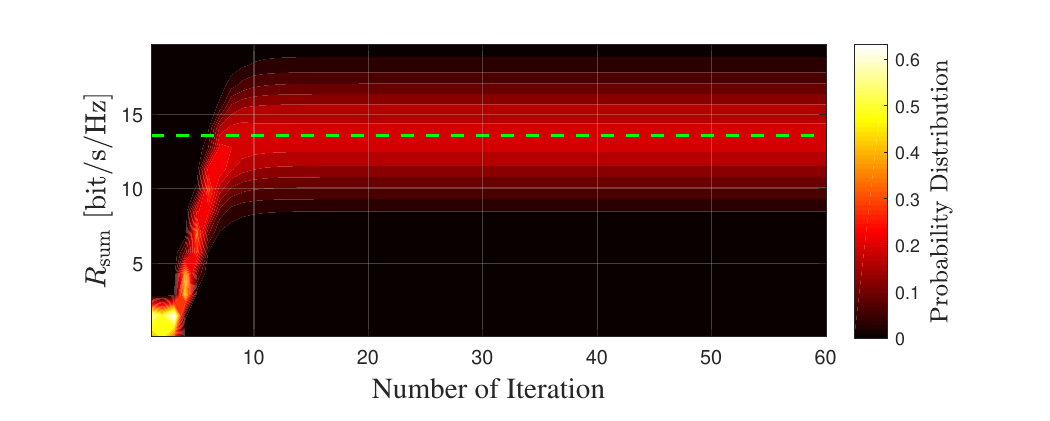}
    \vspace{-3mm}
    \caption{Distribution of sum rate $R_{\text{sum}}$ over iterations.}
    \label{fig:6}
    \vspace{-3mm}
\end{figure}

\begin{figure}[t]
    \centering
    \includegraphics[width=0.8\columnwidth]{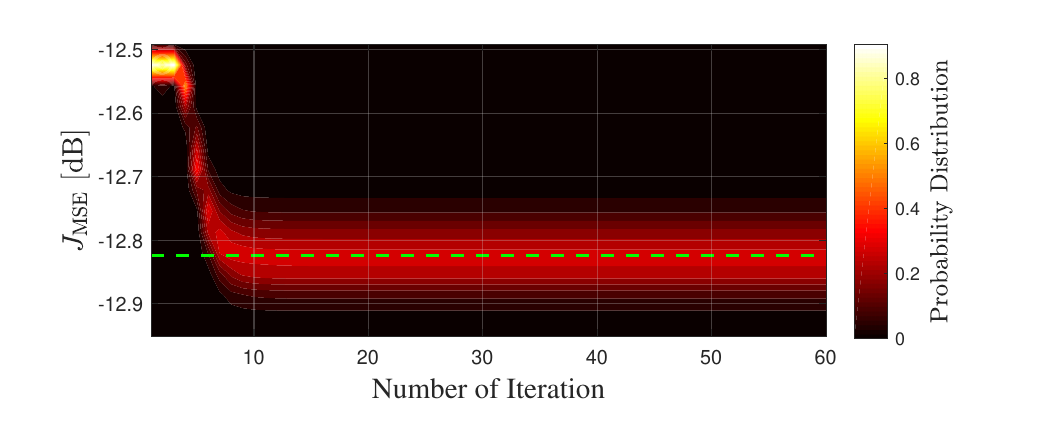}
    \vspace{-3mm}
    \caption{Distribution of beam-matching error $J_{\text{MSE}}$ over iterations.}
    \label{fig:7}
    \vspace{-5mm}
\end{figure}

Figures~\ref{fig:4} and \ref{fig:5} illustrate the effects of weights on the performances of the $\text{D}^3$ algorithm.
Specifically, Fig. \ref{fig:4} illustrates that the beam-matching error diminishes with an increase in $w_1$ and a decrease in $w_2$. Similarly, as shown in Fig. \ref{fig:5}, the sum rate rises with an increase in $w_2$ and a decrease in $w_1$.
Such phenomena are because the adjustment of $w_1$ and $w_2$ can flexibly determine the priority of the normalized gradients of dual objectives, thus being capable of reaching desirable performance trade-offs between sensing and communications under various application requirements.
For scenarios requiring a minimum communication sum rate of $5\,$bit/s/Hz, we identify weight combinations below this threshold in Fig. \ref{fig:4} (illustrated by the red plane) as infeasible, while other weight combinations form the feasible region, corresponding to the gray and colored surfaces in Fig. \ref{fig:5}, respectively. Among the feasible region, the optimal weights $w_1 = 0.6$ and $w_2 = 0.2$ are determined by finding the minimum $J_{\text{MSE}}$ in Fig. \ref{fig:5}, achieving $J_{\text{MSE}} = -13.16\,$dB while maintaining $R_{\text{sum}} = 5\,$bit/s/Hz.
This demonstrates that our algorithm can effectively balance dual-functional performances through systematic weight selection under specific scenarios.

Figures~\ref{fig:6} and \ref{fig:7} present the convergence of the $\text{D}^3$ algorithm, displaying $R_{\text{sum}}$ and $J_{\text{MSE}}$, together with their probability distribution across 100 independent channel realizations, as represented by color intensity. 
Furthermore, the green dashed lines in Figs. \ref{fig:6} and \ref{fig:7} represent the optimal average sum rate and beam-matching error, respectively.
The probability distribution of the convergence trajectories demonstrates that our proposed algorithm consistently converges within approximately $15$ iterations for all channel realizations, thus validating its robust efficacy.
As indicated by the green dashed line in Fig. \ref{fig:6} and Fig. \ref{fig:7}, the average sum rate of our system attains a value of $13.56\,$bit/s/Hz, whereas the average beam-matching error is recorded at $-12.82\,$dB.
Distinguished by the color intensity, the proposed system attains a sum rate above $12\,$bit/s/Hz and a beam-matching error below $-12.8\,$dB in more than $70\%$ of the channel realizations.
Specifically, around $30\%$ of the channel realizations achieve a sum rate between $12$ and $14\,$bit/s/Hz while maintaining a beam-matching error between $-12.84$ and $-12.81\,$dB.
This reveals that our algorithm achieves robust performance against time variations of the channels, confirming its reliability in practical ISAC systems.

\vspace{-3mm}
\section{Conclusion}
%This paper proposed a novel beamforming design for ISAC applications under the framework of transmit SIM-based antenna array.
This paper proposed a novel transmit beamforming design for ISAC applications utilizing SIM architecture, which significantly reduces hardware complexity and power consumption through fully passive wave-domain beamforming.
To achieve balanced dual-function performance trade-off, a non-convex MOP was established by simultaneously maximizing the communication sum rate and optimally shaping the sensing beam pattern by adjusting phase responses of the meta-atoms. To solve this issue, the $\text{D}^3$ algorithm was proposed, which attained satisfactory trade-off between communication and sensing through gradient differences and dual normalization. 
% \textcolor{red}{Numerical results validated that the proposed algorithm can effectively achieve desired trade-offs between communication and sensing performance through appropriate weight adjustment.}
Numerical results confirmed the superiority of our proposed ISAC beamforming design under time-variant channels.

%proposed a novel SIM-based ISAC system design, where SIM is innovatively deployed at the transmitter side to simultaneously perform downlink communication and radar target detection. By solely relying on the phase shifts of SIM, our approach significantly reduces the hardware complexity and cost of the BS while achieving ISAC functionality. 

\vspace{-3mm}
\bibliographystyle{IEEEtran}
\small\bibliography{reference}

\end{document}